\pdfoutput=1 	

\documentclass[preprint,showpacs,preprintnumbers,amsmath,amssymb]{revtex4}

\usepackage{graphicx}
\usepackage{dcolumn}
\usepackage{bm}

\def\lesssim{\ \raise.3ex\hbox{$<$}\kern-0.8em\lower.7ex\hbox{$\sim$}\ }
\def\gesim{\ \raise.3ex\hbox{$>$}\kern-0.8em\lower.7ex\hbox{$\sim$}\ }

\begin{document}
\title{Local Photoemission spectra and effects of spatial inhomogeneity in the BCS-BEC crossover regime of a trapped ultracold Fermi gas}
\author{Miki Ota$^1$, Hiroyuki Tajima$^2$, Ryo Hanai$^2$, Daisuke Inotani$^2$, and Yoji Ohashi$^2$}
\affiliation{$^1$INO-CNR BEC Center and Dipartimento di Fisica, Universit\`a di Trento, 38123 Povo, Italy }
\affiliation{$^2$Department of Physics, Keio University, 3-14-1 Hiyoshi, Kohoku-ku, Yokohama 223-8522, Japan}
\date{\today}
\begin{abstract}
We theoretically investigate single particle excitations in the BCS (Bardeen-Cooper-Schrieffer)-BEC (Bose-Einstein-Condensation) crossover regime of an ultracold Fermi gas. Including strong pairing fluctuations within a $T$-matrix approximation, as well as effects of a harmonic trap potential in the local density approximation, we calculate the local photoemission spectrum in the normal state. Recently, JILA group has measured this quantity in a $^{40}$K Fermi gas, in order to examine {\it homogeneous} single-particle properties of this system. Comparing our results with this experiment, we show that, this attempt indeed succeeds under the JILA's experimental condition. However, we also find that the current local photoemission spectroscopy still has room for improvement, in order to examine the pseudogap phenomenon predicted in the BCS-BEC crossover region. Since ultracold Fermi gases are always in a trap, our results would be useful in applying this system to various {\it homogeneous} Fermi systems, as a quantum simulator.
\end{abstract}
\maketitle
\par
\section{Introduction}
\par
While the high tunability of various physical parameters is an advantage of ultracold Fermi gas\cite{Bloch2008,Giorgini2008,Chin2010,Levin2005}, the spatial inhomogeneity by a trap potential may be a weak point of this many-body system, especially when one tries to use it as a quantum simulator for other {\it uniform} Fermi systems. Using a tunable interaction associated with a Feshbach resonance\cite{Chin2010,Timmermans2001}, we can now study superfluid properties of $^{40}$K and $^6$Li Fermi gases at various interaction strengths in a systematic manner\cite{Regal2004,Zwierlein2004,Bartenstein2004,Kinast2004}. Thus, when the above-mentioned weak point is overcome, one would be able to concentrate on the so-called BCS (Bardeen-Cooper-Schrieffer)-BEC (Bose-Einstein condensation) crossover phenomenon\cite{Levin2005,Nozieres1985,SadeMelo1993,Randeria1995,Ohashi2002,Perali2002,Pieri2004,Ketterle2008,Zwerger2012}, without being annoyed by the unwanted spatial inhomogeneity. Since understanding strong correlations is crucial in various Fermi systems, such as high-$T_{\rm c}$ cuprates\cite{Renner1998,Perali2000,Damascelli2003,Chen2006}, as well as neutron stars\cite{Dean2003}, such improvement would also make an impact on these fields.
\par
So far, several ideas have been proposed/developed to overcome this problem. One idea is to use the combined Gibbs-Duhem equation with the local density approximation (LDA)\cite{Ho2010}, which enables us to evaluate the pressure $P$ of a {\it uniform} Fermi gas from the observed density profile in a {\it trapped} Fermi gas. Then, we can determine other thermodynamic quantities in a uniform Fermi gas from $P$ by way of appropriate thermodynamic identities. This method is now widely used in cold Fermi gas physics\cite{Salomon2010,Zwierlein2012,Horikoshi2016}.
\par
To suppress effects of a harmonic potential, non-harmonic trap potential has also been tested\cite{Hadzibabic2013, Zwierlein2016}. Very recently, for a $^6$Li Fermi gas in a cylindrical trap potential (where atoms feel almost uniform potential in the central region), Zwierlein and co-workers have measured the atomic momentum distribution, to observe a clear signature of the Fermi surface formation below the Fermi temperature $T_{\rm F}$\cite{Zwierlein2016}.
\par
In addition to these, Jin and co-workers have recently invented a local photoemission-type spectroscopy (LPES)\cite{Sagi2015}. While the previous photoemission-type spectroscopy (PES) has no spatial resolution\cite{Stewart2008,Gaebler2010}, LPES employs a space selective imaging technique\cite{Drake2012,Sagi2012,Sagi2013}, to only observe single-particle excitations around the trap center (where effects of a harmonic trap is weak). Since the combined Gibbs-Duhem equation with LDA is only valid for thermodynamic quantities, LPES is expected to contribute to the understanding of {\it excitation} properties of a uniform Fermi gas in the crossover region. 
\par
\begin{figure}[t]
\begin{center}
\includegraphics[keepaspectratio,scale=0.5]{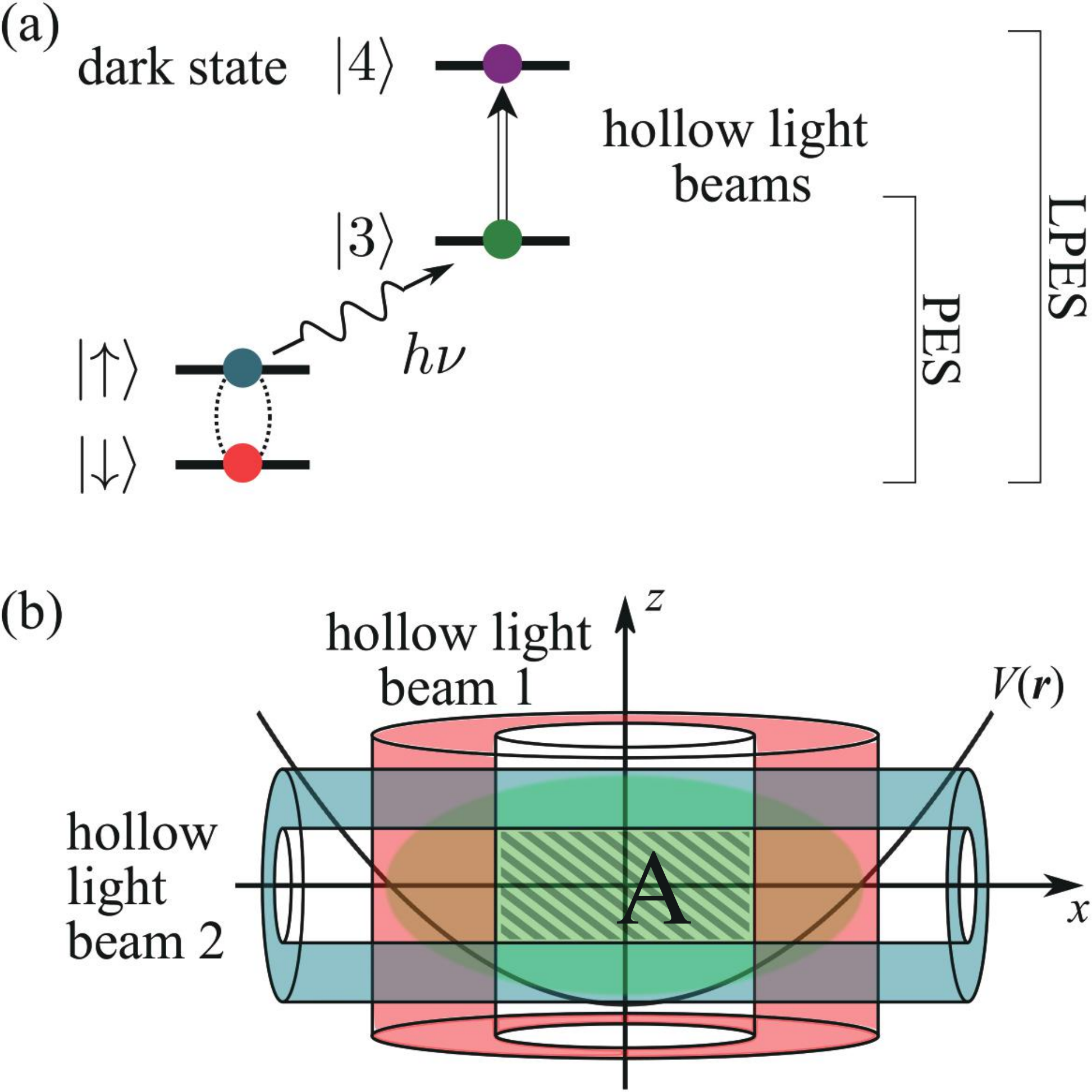}
\caption{Schematic explanation for the local photoemission spectroscopy (LPES). (a) In the ordinary PES, rf-photons are applied to a Fermi gas consisting of two atomic hyperfine states ($|\sigma\rangle=|\uparrow,\downarrow\rangle$), which excite $|\uparrow\rangle$-state atoms to the third state $|3\rangle$. In LPES, after this process, additional hollow light beams are applied to the gas (see panel (b)), to further excite $|3\rangle$-state atoms around the edge of the gas cloud to the fourth state $|4\rangle$. $|4\rangle$ is a dark state in the sense that it is invisible in TOF measurement. This additional manipulation enables LPES to selectively observe single-particle excitations in the central region (``A" in panel (b)).}
\label{Fig1}
\end{center}
\end{figure}
\par
Figure \ref{Fig1} schematically shows difference between PES and LPES. In PES\cite{Stewart2008,Gaebler2010,Torma2000,Torma2016}, rf-photons are applied to the whole gas cloud, to transfer atoms in the $|\uparrow\rangle$-state to the third one $|3\rangle$, as shown in panel (a). The photoemission spectrum is then obtained by probing $|3\rangle$, by using a time-of-flight (TOF) technique. In LPES\cite{Sagi2015}, on the other hand, before TOF measurement, additional hollow light beams are applied to the system (Fig. \ref{Fig1}(b)), to selectively excite $|3\rangle$-state atoms around the edge of the gas cloud to the fourth state $|4\rangle$ (``LPES" in Fig. \ref{Fig1}(a)). When $|4\rangle$ is chosen so as to be invisible in the successive TOF measurement, LPES can selectively probe atoms in the trap center. 
\par
Because of this advantage, LPES is also expected to be able to resolve a long-standing debate on the pseudogap problem in this field. Since pairing fluctuations are strong in the BCS-BEC crossover region, the formation of preformed Cooper pairs, as well as the associated pseudogap phenomenon (where a gap-like structure appears in the single-particle density of states (DOS) in the normal state) have theoretically been predicted\cite{Tsuchiya2009,Watanabe2010,Hui2010,Chien2010,Perali2011,Chen2014}. Experimentally, although the direct observation of DOS is still difficult in cold Fermi gas physics, the recent PES experiment on a $^{40}$K Fermi gas\cite{Stewart2008,Gaebler2010} has found an anomalous spectral structure which is consistent with the prediction\cite{Tsuchiya2009,Watanabe2010,Hui2010,Chien2010,Perali2011,Chen2014}. However, it has also been argued that the observed anomaly is not a signature of pseudogap, but simply comes from the spatial inhomogeneity of a trapped Fermi gas\cite{Nascimbene2011,Nascimbene2010}, so that the existence of this many-body phenomenon is still controversial. To resolve this debate, LPES is very promising, because this experiment can, in principle, eliminate effects of a trap from the spectrum. Since the pseudogap phenomenon has also been discussed in high-$T_{\rm c}$ cuprates as a key to clarify the pairing mechanism\cite{Perali2002,Randeria1992,Singer1996,Janko1997,Yanase2001,Rohe2001}, the observation of pseudogap in the simpler  gas system would also contribute to the study of this more complicated electron system. 
\par
In the recent LPES experiment on a $^{40}$K Fermi gas\cite{Sagi2015}, the observed spectrum is similar to the previous PES result\cite{Stewart2008,Gaebler2010}, which looks supporting the pseudogap scenario. However, because the current LPES\cite{Sagi2015} still needs at least about 30\% of trapped atoms in order to obtain detectable spectral intensity, it must be done, not at the trap center, but for a finite volume fraction around the trap center. Thus, it is still unclear whether the observed LPES spectra really describe single particle properties of a {\it homogeneous} system or they still involve non-negligible inhomogeneous effects by a harmonic trap. 
\par
The purpose of this paper is to theoretically investigate the local photoemission-type spectroscopy (LPES) in the BCS-BEC crossover regime of a trapped normal Fermi gas. Including pairing fluctuations in the BCS-BEC crossover region within the framework of a strong-coupling $T$-matrix approximation (TMA), as well as effects of a harmonic trap within the local density approximation (LDA), we assess to what extent LPES can eliminate effects of a harmonic trap from the photoemission spectrum. We clarify the region where the current LPES technique can obtain single-particle properties of a {\it uniform} Fermi gas, in the phase diagram with respect to the temperature and the strength of pairing interaction. We also discuss how the local photoemission spectrum is sensitive to detailed spatial selection.  
\par
This paper is organized as follows. In Sec. II, we explain our formulation. In addition to the combined TMA with LDA to describe a trapped Fermi gas in the BCS-BEC crossover region, we also explain how to theoretically evaluate the local photoemission spectrum. In Sec. III, we show our numerical results. We compare calculated spectra with the recent experiment on a $^{40}$K Fermi gas\cite{Sagi2015}. Throughout this paper, we set $\hbar=k_{\rm B}=1$, for simplicity. In the uniform case, the system volume $V$ is taken to be unity.
\par
\section{Formulation}
\par
We start from a two-component Fermi gas described by the BCS Hamiltonian, 
\begin{equation}
H=\sum_{{\bm p},\sigma} \xi_{\bm p} c_{{\bm p},\sigma}^\dagger c_{{\bm p},\sigma} - U \sum_{{\bm p},{\bm p}',{\bm q}} c_{{\bm p}+{\bm q},\uparrow}^\dagger c_{{\bm p}'-{\bm q},\downarrow}^\dagger c_{{\bm p}',\downarrow} c_{{\bm p},\uparrow}.\label{eq.1}
\end{equation}
We first explain TMA formalism and PES expression in the {\it uniform} case. Later, we explain how to incorporate effects of a harmonic trap into the theory within LDA. In Eq. (\ref{eq.1}), $c^\dagger_{{\bm p},\sigma}$ is the creation operator of a Fermi atom in the hyperfine state described by pseudo-spin $\sigma=\uparrow,\downarrow$. $\xi_{\bm p}=\varepsilon_{\bm p}-\mu={\bm p}^2/(2m)-\mu$ is the kinetic energy, measured from the Fermi chemical potential $\mu$, where $m$ is an atomic mass. $-U~(<0)$ is an attractive pairing interaction, which is assumed to be tunable by a Feshbach resonance technique\cite{Chin2010}. As usual, we measure the interaction strength in terms of the observable $s$-wave scattering length $a_{\rm s}$, which is related to $-U$ as,
\begin{equation}
{4\pi a_s \over m}=
-{U \over 1-U\sum_{\bm p}^{p_{\rm c}}1/(2\varepsilon_{\bm p})}.
\label{eq.2}
\end{equation}
where $p_{\rm c}$ is a momentum cutoff. 
\par
\begin{figure}[t]
\begin{center}
\includegraphics[width=0.45\linewidth]{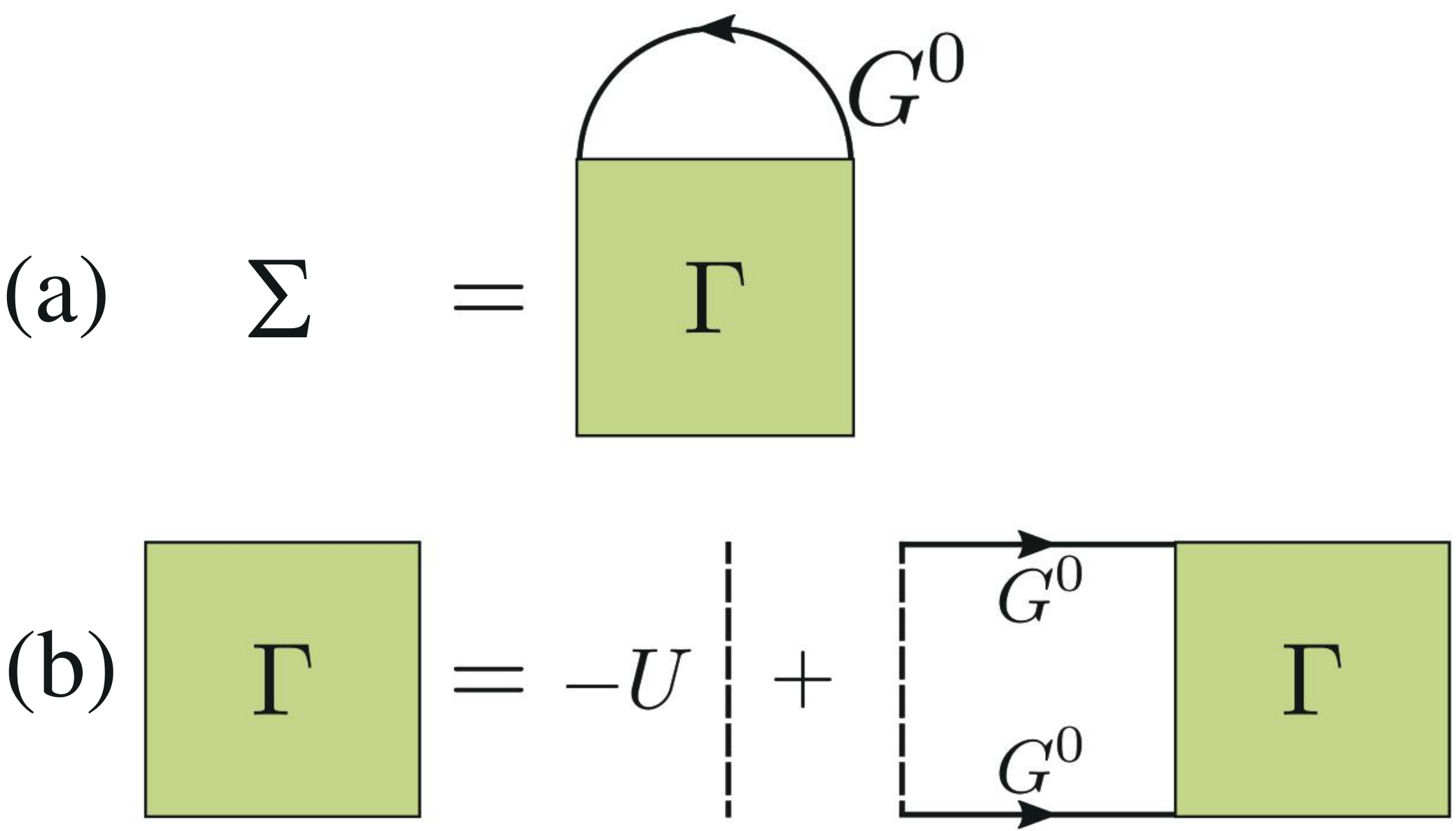}
\caption{(a) TMA self-energy $\Sigma({\bm p},i\omega_n)$. The particle-particle scattering matrix $\Gamma({\bm q},i\nu_n)$ is given by the sum of the ladder diagrams shown in panel (b). $G^0=1/[i\omega_n-\xi_{\bm p}]$ is the bare single-particle Green's function. $-U~(<0)$ is a pairing interaction. }
\label{Fig2}
\end{center}
\end{figure}
\par
Strong-coupling corrections to single-particle excitations are conveniently described by the self-energy $\Sigma({\bm p},i\omega_n)$ in the single-particle thermal Green's function\cite{Mahan1981,Rickayzen2013},
\begin{equation}
G({\bf p},i\omega_n)=
{1 \over i\omega_n-\xi_{\bm p}-\Sigma({\bm p},i \omega_n)},
\label{eq.3}
\end{equation}
where $\omega_n$ is the fermion Matsubara frequency. In this paper, we deal with the self-energy correction $\Sigma({\bm p},i\omega_n)$ within the $T$-matrix approximation (TMA)\cite{Perali2002,Pieri2004,Tsuchiya2009,Watanabe2010}, which is diagrammatically given in Fig. \ref{Fig2}. We briefly note that TMA can describe the BCS-BEC crossover behavior of $T_{\rm c}$, as well as the pseudogapped density of states in the crossover region\cite{Perali2002,Pieri2004,Tsuchiya2009,Watanabe2010}. This strong-coupling theory has also been used to calculate the conventional PES\cite{Tsuchiya2011}, to successfully explain the observed spectra in a $^{40}$K Fermi gas\cite{Stewart2008,Gaebler2010}. Thus, we expect that TMA is also suitable for our purpose.
\par
The summation of the TMA diagrams in Fig. \ref{Fig2} gives
\begin{equation}
\Sigma({\bm p},i\omega_n)=T\sum_{{\bm q},i\nu_n}
\Gamma({\bm q},i\nu_n)G^0 ({\bm q}-{\bm p},i\nu_n-i\omega_n).
\label{eq.3b}
\end{equation}
Here, $i\nu_n$ is the boson Matsubara frequency, and $G^{0}({\bm p},i\omega_n)= 1/[i\omega_n-\xi_{\bm p}]$ is the bare single-particle Green's function. In Eq. (\ref{eq.3b}), the particle-particle scattering matrix,
\begin{equation}
\Gamma({\bm q},i\nu_n)=
-{U \over 1-U\Pi({\bm q},i\nu_n)},
\label{eq.4}
\end{equation}
describes fluctuations in the Cooper channel, where
\begin{eqnarray}
\Pi({\bm q},i\nu_n)
&=&
T\sum_{{\bm p},\omega_n}
G^0({\bm p}+{\bm q}/2,i\omega_n+i\nu_n)
G^0(-{\bm p}+{\bm q}/2,-i\omega_n)
\nonumber
\\
&=&
\sum_{\bm p}
{1-f(\xi_{{\bm p}+{\bm q}/2})-f(\xi_{-{\bm p}+{\bm q}/2})
\over 
\xi_{{\bm p}+{\bm q}/2}+\xi_{-{\bm p}+{\bm q}/2}-i\nu_n}
\label{eq.5}
\end{eqnarray}
is the lowest-order pair correlation function, with $f(x)$ being the Fermi distribution function. As usual, the momentum summation in Eq. (\ref{eq.5}) involves the ultraviolet divergence, which however, does not affect the final result, when the interaction strength is measured in terms of the $s$-wave scattering length $a_s$ as,
\begin{equation}
\Gamma({\bm q},i\nu_n)=
{\displaystyle
{4\pi a_s \over m} 
\over 
\displaystyle
1+{4\pi a_s \over m}
\left[\Pi({\bm q},i\nu_n)-\sum_{\bm p}{1 \over 2\varepsilon_{\bm p}}\right]
}.
\label{eq.6}
\end{equation}
\par
To describe excitations from $|\uparrow\rangle$ to $|3\rangle$ by rf-photons\cite{Stewart2008,Gaebler2010}, we add the corresponding Hamiltonian\cite{Torma2000,Bruun2001,Kinnuen2004,Yan2005,Chen2009},
\begin{equation}
H_3=
\sum_{\bm p}
\left[\varepsilon_{\bm p}+\omega_3-\mu_3 \right]
b_{\bm p}^\dagger b_{\bm p}
+
t_{\rm rf} \sum_{\bm p}
\left[ e^{-i\omega_{\rm L} t} 
b_{{\bm p}+{\bm q}_{\rm L}}^\dagger
c_{{\bm p},\uparrow}+{\rm h.c.}
\right],
\label{eq.7}
\end{equation}
to our model in Eq. (\ref{eq.1}). In Eq. (\ref{eq.7}), $b_{\bm p}^\dagger$ is the creation operator of a Fermi atom in $|3\rangle$ with the kinetic energy $\varepsilon_{\bm p}+\omega_3-\mu_3$, measured from the chemical potential $\mu_3$ (where $\omega_3$ is the energy difference between $|\uparrow\rangle$ and $|3\rangle$). The last term in Eq. (\ref{eq.7}) ($\equiv H_{\rm T}$) describes a photon-assisted tunneling between $|\uparrow\rangle$ and $|3\rangle$ in the rotational wave approximation\cite{Torma2000,Bruun2001,Kinnuen2004,Yan2005,Chen2009,Ohashi2005}, where $t_{\rm rf}$ is the transfer matrix element between the two states, and ${\bm q}_{\rm L}$ and $\omega_{\rm L}$ are the momentum and energy of rf-photon, respectively. 
\par
The tunneling current $I({\bm p},t)$ from $|\uparrow\rangle$ to $|3\rangle$ is given as the increase rate of the number $N_3({\bm p},t)$ of $|3\rangle$-state atoms with momentum ${\bm p}$. Treating the tunneling Hamiltonian $H_{\rm T}$ within the linear response theory\cite{Mahan1981}, one obtains
\begin{equation}
I({\bm p},t)=\langle \dot{N}_3({\bm p},t)\rangle=
i\int_{-\infty}^t {\rm d}t' 
\langle[ H_{\rm T}(t') ,\hat{I}({\bm p},t)]\rangle,
\label{eq.8}
\end{equation}
where 
\begin{equation}
\hat{I}({\bm p},t) = \dot{N}_3({\bm p},t)=-i t_{\rm rf} \left[ e^{-i\omega_{\rm L} t} b_{{\bm p}+{\bm q}_{\rm L}}^\dagger(t) c_{{\bm p},\uparrow}(t)-{\rm h.c.} \right] 
\end{equation}
is a current operator, with $H_{\rm T}(t')=e^{iHt'}H_{\rm T}e^{-iHt'}$. The photoemission spectrum $I({\bm p},\omega)$ is given as the Fourier transformed tunneling current $I({\bm p},t)$ in terms of the time-variable $t$. Assuming that the photon momentum ${\bm q}_{\rm L}$ is negligibly small and the third state $|3\rangle$ is initially vacant ($f(\varepsilon_{\bm p}+\omega_3-\mu_3)=0$), we have,
\begin{equation}
I({\bm p},\Omega)=2\pi t_{\rm rf}^2A({\bm p},\xi_{\bm p}-\Omega)
f(\xi_{\bm p}-\Omega).
\label{eq.9}
\end{equation}
Here, $\Omega\equiv \omega_3+\mu_3-\mu-\omega_{\rm L}$ is sometimes referred to as the detuning frequency in the literature. Equation (\ref{eq.9}) involves the single-particle spectral weight $A({\bm p},\omega)$, which is related to the analytic continued single-particle Green's function in Eq. (\ref{eq.3}) as
\begin{equation}
A({\bm p},\omega) = -{1 \over \pi}{\rm Im}
\left[G({\bm p},i\omega_n\rightarrow \omega + i \delta)\right],
\label{eq.10}
\end{equation}
where $\delta$ is an infinitesimally small positive number. Thus, the photoemission spectrum $I({\bm p},\Omega)$ in Eq. (\ref{eq.9}) gives us useful information about many-body corrections to single-particle excitations in the BCS-BEC crossover region.
\par
We now include effects of a harmonic trap. In LDA, this extension is achieved by simply replacing $\mu$ and $\mu_3$ by the position-dependent ones, $\mu({\bf r}) = \mu - V({\bf r})$ and $\mu_3 ({\bf r}) = \mu_3 - V({\bf r})$, respectively\cite{Tsuchiya2011,Butts1997,Ohashi2003,note1}. Here,
\begin{equation}
V({\bm r})={1 \over 2} m\omega_{\rm tr}^2 r^2
\label{eq.11}
\end{equation}
is a harmonic potential with the trap frequency $\omega_{\rm tr}$\cite{note2}. Equations (\ref{eq.3})-(\ref{eq.6}) then depend on ${\bm r}$ through $\mu({\bm r})$. For example, the LDA single-particle thermal Green's function has the form,
\begin{equation}
G({\bm p},i\omega_n,{\bm r})={1 \over i\omega_n-\xi_{\bm p}({\bm r})-\Sigma({\bm p},i\omega_n,{\bm r})},
\label{eq.12}
\end{equation}
where $\xi_{\bm p}({\bm r})=\varepsilon_{\bm p}-\mu({\bm r})$. The LDA photoemission spectrum is given by,
\begin{equation}
I({\bm p},\Omega,{\bm r})=2\pi t_{\rm rf}^2
A({\bm p},\xi_{\bm p}({\bm r})-\Omega,{\bm r})f(\xi_{\bm p}({\bm r})-\Omega),
\label{eq.13}
\end{equation}
where $A({\bm p},\omega,{\bm r})=-{\rm Im}[G({\bm p},i\omega\to\omega+i\delta,{\bm r})]/\pi$. We briefly note that $\Omega$ in Eq. (\ref{eq.13}) is still ${\bm r}$-independent in LDA. 
\par
In the conventional photoemission spectroscopy (PES) with no spatial resolution\cite{Stewart2008,Gaebler2010}, the spectrum $I_{\rm PES}({\bm p},\omega)$ is obtained from the spatial average of Eq. (\ref{eq.13}) over the {\it entire} gas cloud. When we slightly modify the expression so that we can directly compare our results with experimental data, we have 
\begin{equation}
I_{\rm PES}({\bm p},\omega)={2\pi t_{\rm rf}^2 \over V}p^2
\int {\rm d}{\bm r}A({\bm p},\omega-\mu({\bm r}),{\bm r})
f(\omega-\mu({\bm r})).
\label{eq.14}
\end{equation} 
Here, $V=4\pi R_{\rm F}^3/3$ is a characteristic volume of the gas cloud, where  $R_{\rm F}=(24Nk_{\rm F}^{-3})^{1/3}$ is the Thomas-Fermi radius\cite{Pethick}  (where $N$ and $k_{\rm F}$ are the number of Fermi atoms and the Fermi momentum in the trap center in LDA, respectively). In obtaining Eq. (\ref{eq.14}), we have changed the variable as $\Omega=\xi_{\bm p}-\omega$, and have multiplied the spectrum by $p^2$, following the PES experiment\cite{Stewart2008,Gaebler2010}. 
\par
\begin{figure}[t]
\begin{center}
\includegraphics[width=0.5\textwidth]{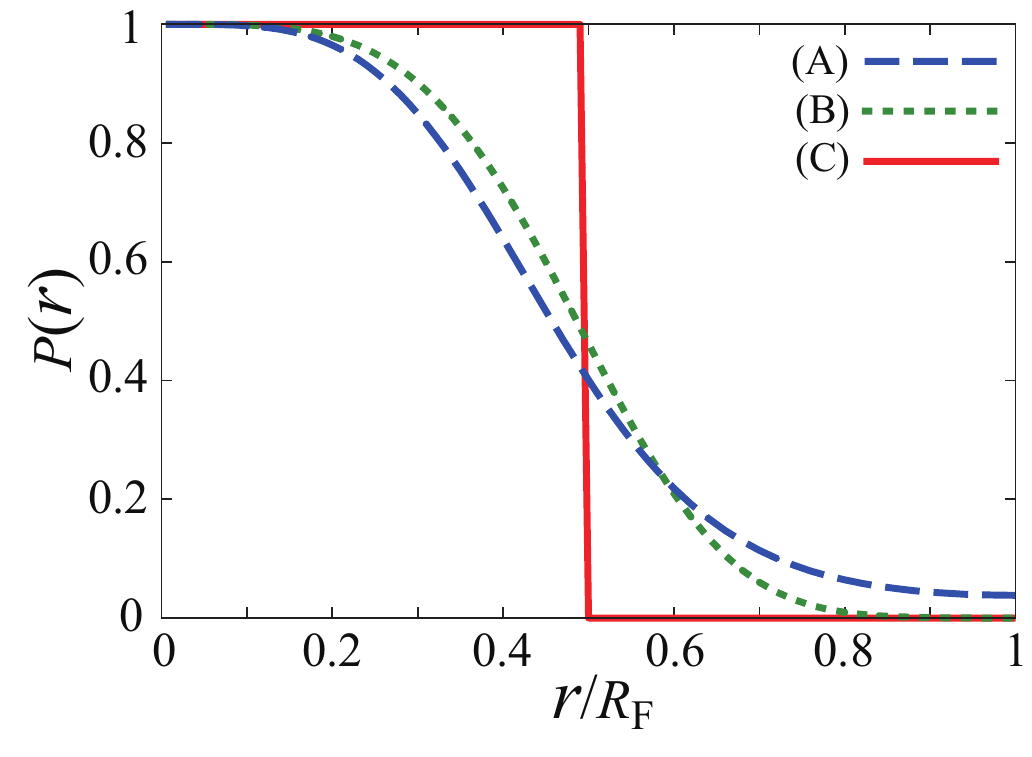}
\caption{Space selective function $P(r)$ in Eq. (\ref{eq.16}), as a function of $r$. (A) $w/R_{\rm F}=1$, and $J_0/R_{\rm F}^2 = 6$. This parameter set is used in Fig. \ref{Fig4}(a). (B) $w/R_{\rm F}=3$ and $J_0/R_{\rm F}^2 = 2380$. (C) shows the the case with sharp cutoff in Eq. (\ref{eq.18}) with $R_{\rm cut}/R_{\rm F}=0.48$. This case is examined in Fig. \ref{Fig4}(c). In all the three cases, when one considers the case shown in Fig. \ref{Fig4}, 30\% of trapped atoms contribute to the local photoemission spectrum.}
\label{Fig3}
\end{center}
\end{figure} 
\par
As mentioned in Sec. I, the local photoemission spectroscopy (LPES) developed by JILA group\cite{Sagi2015} uses a space selective imaging technique\cite{Drake2012,Sagi2012,Sagi2013}. This can be conveniently incorporated into the theory by inserting a ``space selective function" $P({\bm r})$, which describes the probability that a $|3\rangle$-state atom at ${\bm r}$ is {\it not} scattered into $|4\rangle$, into the PES expression in Eq. (\ref{eq.14}). The resulting LPES spectrum is given by,
\begin{equation}
I_{\rm LPES}({\bm p},\omega) ={2\pi t_{\rm rf}^2 \over V}p^2
\int {\rm d}{\bm r} A({\bm p},\omega-\mu({\bm r}),{\bm r})f(\omega-\mu({\bm r})) P({\bm r}).
\label{eq.15}
\end{equation}
In the JILA's experiment\cite{Sagi2015}, the space selection has been done by using two hollow light beams, both of which have the second-order Laguerre-Gaussian profile\cite{Drake2012}, to kick out atoms in the outer region of the gas cloud to the invisible $|4\rangle$-state (see Fig. \ref{Fig1}). To model this set-up, we take\cite{note5}
\begin{equation}
P ({\bm r}) = e^{-{J(r) \over w}},
\label{eq.16}
\end{equation}
where
\begin{equation}
J(r) = J_0\left({2r^2 \over w^2}\right)^2 e^{-{2r^2 \over w^2}},
\label{eq.17}
\end{equation}
with $J_0$ and $w$ describing the power and the waist of the beam, respectively. For clarity, we show the detailed spatial variation of $P({\bm r})$ in Fig. \ref{Fig3}. We briefly note that, although $P({\bm r})$ in Eq. (\ref{eq.16}) does not exactly reproduce the experimental set-up (with two hollow light beams, see Fig. \ref{Fig1}(b))\cite{Sagi2015,note5}, we will later show that the LPES spectrum is actually not so sensitive to the detailed spatial variation of $P({\bm r})$. 
\par
In this paper, we also consider the case with a {\it sharp} cutoff,
\begin{equation}
P({\bm r})=\Theta(R_{\rm cut}-r),
\label{eq.18}
\end{equation}
where $\Theta(x)$ is the step function, and $R_{\rm cut}$ is a cutoff radius.
\par
Before ending this section, we comment on our numerical calculations. We first determine the Fermi chemical potential $\mu(T)$ in the normal state from the LDA  number equation\cite{Ohashi2003},
\begin{equation}
N=2T\int {\rm d}{\bm r}\sum_{{\bm p},i\omega_n}e^{i\omega_n \delta}
G({\bm p},i\omega_n,{\bm r}).
\label{eq.19}
\end{equation}
Using the calculated $\mu(T)$, we evaluate the local photoemission spectrum $I_{\rm LPES}({\bm p},\omega)$ in Eq. (\ref{eq.15}). For this purpose, we numerically execute the analytic continuation, $G({\bm p},i\omega_n\to\omega+i\delta,{\bm r})$, in the Pad\'e approximation\cite{Vidberg1977,Beach2000}, to evaluate the spectral weight $A({\bm p},\omega,{\bm r})$ in Eq. (\ref{eq.15}). In this paper, we focus on the normal state above the superfluid phase transition temperature $T_{\rm c}$. In LDA, $T_{\rm c}$ is determined as the temperature at which the Thouless criterion\cite{Thouless1960} is satisfied at the trap center (${\bm r}=0$)\cite{Ohashi2003}, that is,
\begin{equation}
\Gamma({\bm q}={\bm 0},i\nu_n=0,{\bm r}={\bm 0}) ^{-1} = 0.
\label{eq.20}
\end{equation}  
This condition gives the same form as the ordinary $T_{\rm c}$-equation in the BCS theory as,
\begin{equation}
1=-{4\pi a_s \over m}\sum_{\bm p}
\left[
{1 \over 2\xi_{\bm p}}\tanh{\xi_{\bm p} \over 2T_{\rm c}}-{1 \over 2\varepsilon_{\bm p}}
\right].
\label{eq.21}
\end{equation}
As usual, we numerically solve Eq. (\ref{eq.21}) together with the number equation (\ref{eq.19}), to self-consistently determine $T_{\rm c}$ and $\mu(T_{\rm c})$. 
\par
\begin{figure}[t]
\begin{center}
\includegraphics[width=0.6\textwidth]{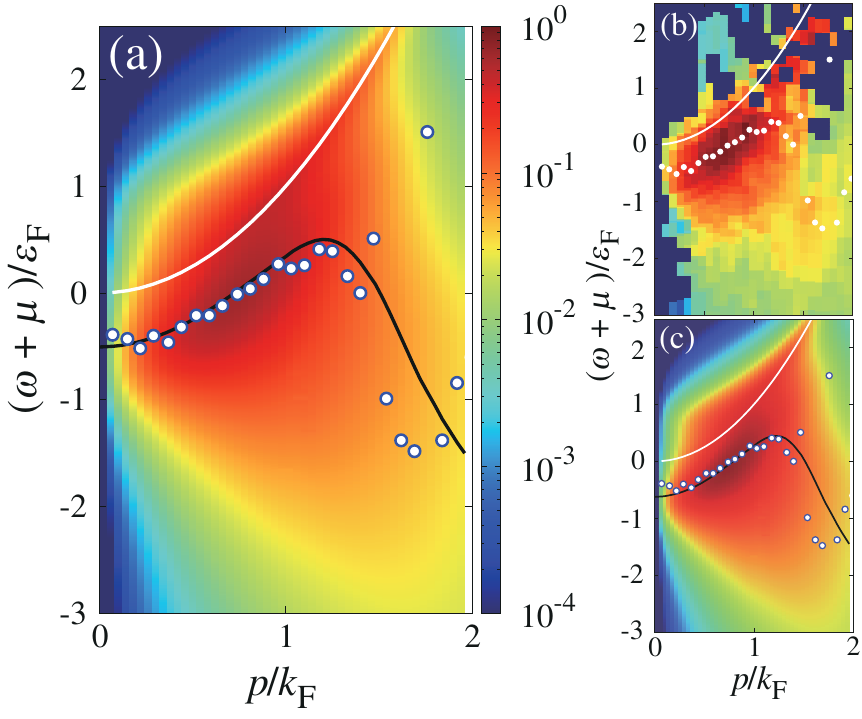}
\caption{Comparison of calculated intensity $I_{\rm LPES}({\bm p},\omega)$ of the local photoemission spectrum (a) with the recent experiment on a $^{40}$K Fermi gas (b)\cite{Sagi2015}. Both are in the unitary regime ($(k_{\rm F} a_s)^{-1} = 0.1$) at $T/T_{\rm c}= 1.24$, and 30\% of atoms in the trap center contribute to the spectrum. To realize this situation, we set $w/R_{\rm F}=1$, and $J_0/R_{\rm F}^2 = 6$ in panel (a) ((A) in Fig. \ref{Fig3}). (c) Calculated spectral intensity in the case of sharp cutoff in Eq. (\ref{eq.18}). To reproduce the experimental probing rate (30\%), we set $R_{\rm cut}/R_{\rm F}=0.48$ ((C) in Fig. \ref{Fig3}). The other parameters are the same as those in panels (a) and (b). In these figures, the black solid lines and while circles show the peak positions of the calculated and observed spectral intensity, respectively. The white solid line shows the free-particle dispersion, $\omega+\mu={\bm p}^2/(2m)$. The spectral intensity is normalized by $(k_{\rm F}\varepsilon_{\rm F})^{-1}\int_{-\mu-5\varepsilon_{\rm F}}^{-\mu+4\varepsilon_{\rm F}}{\rm d}\omega\int_0^{2k_{\rm F}}{\rm d}p I_{\rm LPES}({\bm p},\omega)$, where $\varepsilon_{\rm F}$ and $k_{\rm F}$ are the Fermi energy and the Fermi momentum in LDA, respectively.
}
\label{Fig4}
\end{center}
\end{figure} 
\par
\section{Local Photoemission spectra in the BCS-BEC crossover region}
\par
Figure \ref{Fig4}(a) shows the calculated intensity $I_{\rm LPES}({\bm p},\omega)$ of the local photoemission spectrum in the unitary regime of an ultracold Fermi gas at $T=1.24T_{\rm c}$. In this figure, $30\%$ of trapped atoms in the central region are selectively probed. We find that the overall structure of the spectral intensity $I_{\rm LPES}({\bm p},\omega)$ agrees well with the recent LPES experiment on a $^{40}$K Fermi gas (Fig. \ref{Fig4}(b))\cite{Sagi2015}. In particular, our result quantitatively explains the back-bending behaviour of the observed spectral peak (white circles in panels (a) and (b)), see the black solid line in Fig. \ref{Fig4}(a). Such agreements are also obtained at different interaction strengths in the crossover region, although we do not explicitly show them in this paper. These agreements confirm the validity of our approach in this regime.
\par
\begin{figure}[t]
\begin{center}
\includegraphics[width=0.5\textwidth]{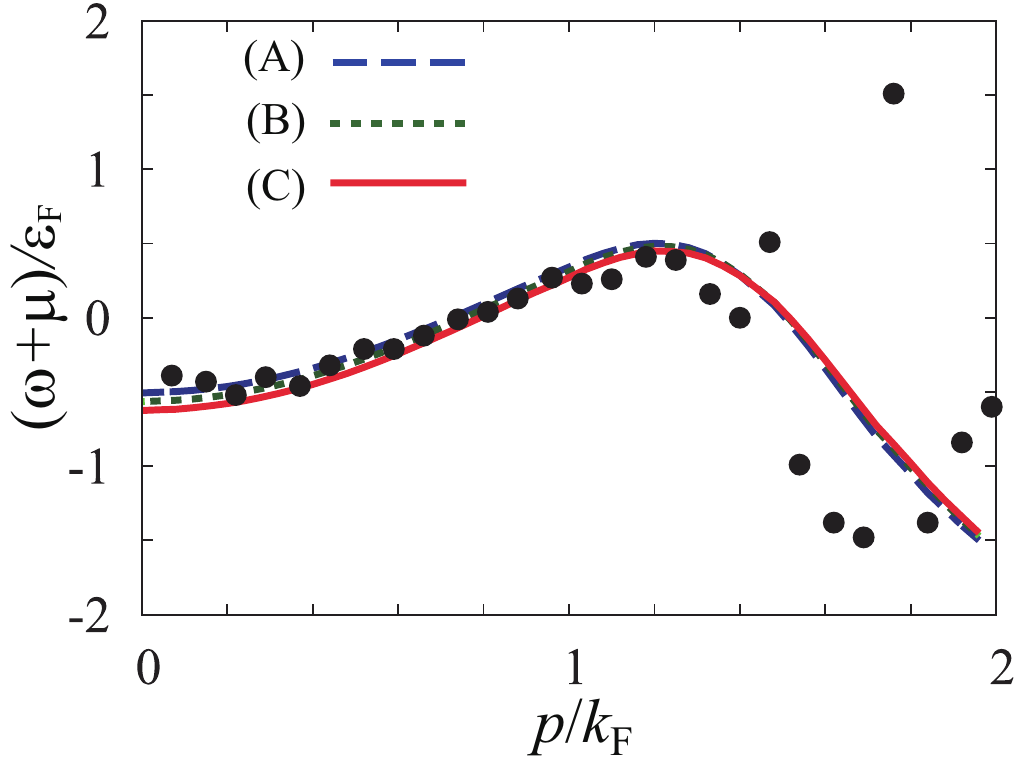}
\caption{Calculated spectral peak lines in the cases (A)-(C) in Fig. \ref{Fig3}. We take $(k_{\rm F} a_s)^{-1} = 0.1$, $T=1.24T_{\rm c}$, and $N_{\rm prob}/N=0.3$. (A) and (C) are the same cases as Figs. \ref{Fig4}(a) and (c), respectively. Solid circles are experimental data\cite{Sagi2015}.}
\label{Fig5}
\end{center}
\end{figure}

Figure \ref{Fig4}(c) shows the case when 30\% of atoms are selected by the sharp cutoff function in Eq. (\ref{eq.18}), which gives almost the same result as Fig. \ref{Fig4}(a). We also see in Fig. \ref{Fig5} that, although the parameter set $(w,J_0)$ in $P({\bm r})$ in Eq. (\ref{eq.16}) which probes 30\% of atoms is not unique, the spectral peak line is not sensitive to this ambiguity. These indicate that, once the probing rate ($\equiv N_{\rm prob}/N$) is fixed, the detailed spatial selection is not so crucial for the local photoemission spectrum. 
\par
\par
To assess to what extent LPES detects single-particle properties of a uniform Fermi gas from a trapped Fermi gas, we conveniently introduce the quantity,
\begin{equation}
h= \sum_{\bm p}\int {\rm d}\omega
\langle\langle I_{\rm uniform} ({\bm p},\omega)/p^2 \rangle\rangle 
\langle\langle I_{\rm LPES} ({\bm p},\omega)/p^2 \rangle\rangle,
\label{eq.22}
\end{equation}
where $I_{\rm uniform} ({\bm p},\omega)$ is the photoemission spectrum in a uniform Fermi gas with the same temperature and interaction strength as the trapped case. The atomic number density in this uniform case is chosen to be equal to the central density in the trapped case. In the present formalism, $I_{\rm uniform} ({\bm p},\omega)$ is given by, 
\begin{equation}
I_{\rm uniform}({\bm p},\omega) =2\pi t_{\rm rf}^2 p^2
A({\bm k},\omega-\mu,{\bm r}=0)f(\omega-\mu).
\label{eq.23}
\end{equation}
In Eq. \eqref{eq.22}, $\langle\langle X({\bm p},\omega) \rangle\rangle$ means the normalization,
\begin{equation}
\langle\langle X({\bm p},\omega) \rangle\rangle=
{X({\bm p},\omega) \over
\sqrt{\sum_{\bm p}\int {\rm d}\omega X^2({\bm p},\omega)}}.
\label{eq.24}
\end{equation}
We note that one obtains $h=1$, when the local photoemission spectrum $I_{\rm LPES}({\bm p},\omega)$ is the same as the uniform result, $I_{\rm uniform}({\bm p},\omega)$. The decrease from the unity ($h<1$) means that the spectrum is still influenced by spatial inhomogeneity by a harmonic trap.
\par
\begin{figure}[t]
\begin{center}
\includegraphics[width=\textwidth]{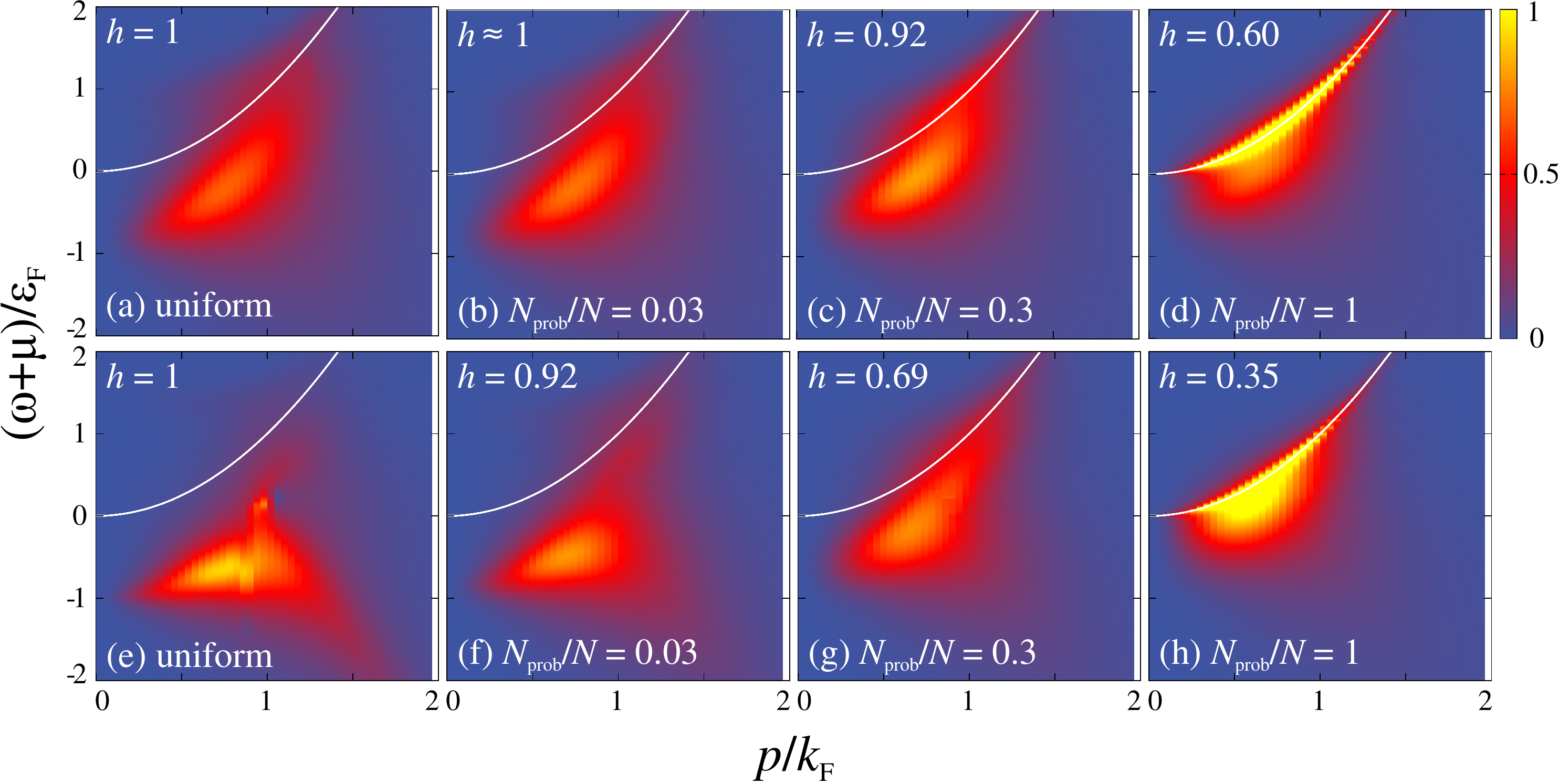}
\caption{(a)-(d) Calculated local photoemission spectra, when $(k_{\rm F} a_s)^{-1}=0.1$ and $T=1.24T_{\rm c}$. (a) Uniform case. (b) $N_{\rm prob}/N=0.02$.  (c) $N_{\rm prob}/N=0.3$. (d) $N_{\rm prob}/N=1$. The case (d) is the same as the conventional photoemission spectrum $I_{\rm PES}({\bm p},\omega)$, where all the atoms in the trap contribute to the spectrum. (e)-(h) Same plots as (a)-(d), but in the unitarity limit ($(k_{\rm F}a_s)^{-1}=0$) at $T_{\rm c}$. The white solid line shows the free-particle dispersion.}
\label{Fig6}
\end{center}
\end{figure}
\par
Figures \ref{Fig6}(a)-(d) show how the probing rate $N_{\rm prob}/N$ affects $I_{\rm LPES}({\bm p},\omega)$, when $(k_{\rm F}a_s)^{-1}=0.1$ and $T=1.24T_{\rm c}$. The spectral structure in the experimental situation (panel (c)) looks closer to the uniform result in panel (a) than the conventional PES case in panel (d) (where all the atoms in the trap contribute to the spectrum). Indeed, the value $h=0.92$ in the case of Fig. \ref{Fig6}(c) is very close to unity. In this sense, the space selective imaging technique in the recent LPES experiment on a $^{40}$K Fermi gas\cite{Sagi2015} is considered to succeed in observing single-particle excitations in a (nearly) uniform Fermi gas.
\par
\begin{figure}[t]
\begin{center}
\includegraphics[width=0.5\textwidth]{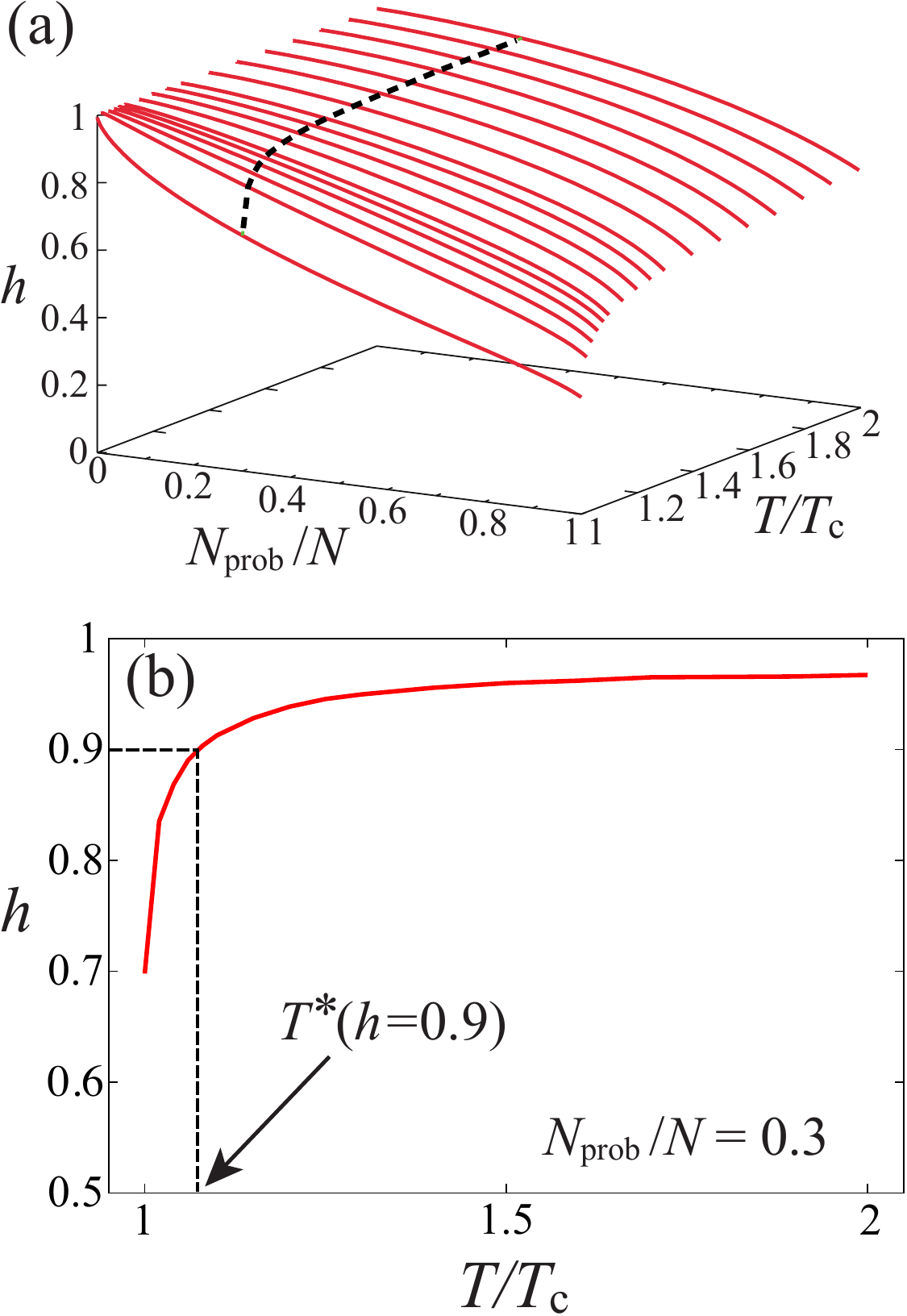}
\caption{(a) Calculated $h$-factor in Eq. (\ref{eq.22}), as a function of the temperature $T$ and the probing rate $N_{\rm prob}/N$. We take $(k_{\rm F}a_s)^{-1}=0$. The dotted line shows the result at $N_{\rm prob}/N=0.3$, which is also shown in the two-dimensional panel (b), for clarity. In panel (b), $T^*(h=0.9)$ is the temperature above which $h\ge 0.9$ is realized. In this figure, the lowest temperature equals $T_{\rm c}$.} 
\label{Fig7}
\end{center}
\end{figure}
\par
However, we point out that the current experimental limitation $N_{\rm prob}/N\gesim 0.3$ is not always enough to obtain single-particle properties of a {\it uniform} Fermi gas. For example, in the unitarity limit at $T_{\rm c}$, Fig. \ref{Fig6}(g) shows that the similarity of the spectrum to the uniform result is at most $h=0.69\ll 1$. Indeed, we see in Fig. \ref{Fig6}(g) that the spectrum has large intensity around the the free-particle dispersion (white solid line), compared to the uniform case in panel (e). This structure is rather close to the PES case shown in panel (h). To obtain $h=0.92$ in this case, one needs to tune the probing rate down to $N_{\rm prob}/N=0.03$ (Fig. \ref{Fig6}(f)), which is, however, beyond the current experimental limitation. 
\par
These results indicate that, not only the probing rate $N_{\rm prob}/N$, but also the temperature is crucial to obtain $h\simeq 1$. Indeed, Fig. \ref{Fig7} shows that the $h$-factor remarkably decreases near $T_{\rm c}$. Thus, one needs to take a smaller value of the probing rate than $N_{\rm prob}/N=0.3$, to obtain single-particle properties of a uniform Fermi gas at $T\simeq T_{\rm c}$.
\par
\begin{figure}[t]
\begin{center}
\includegraphics[width=0.5\textwidth]{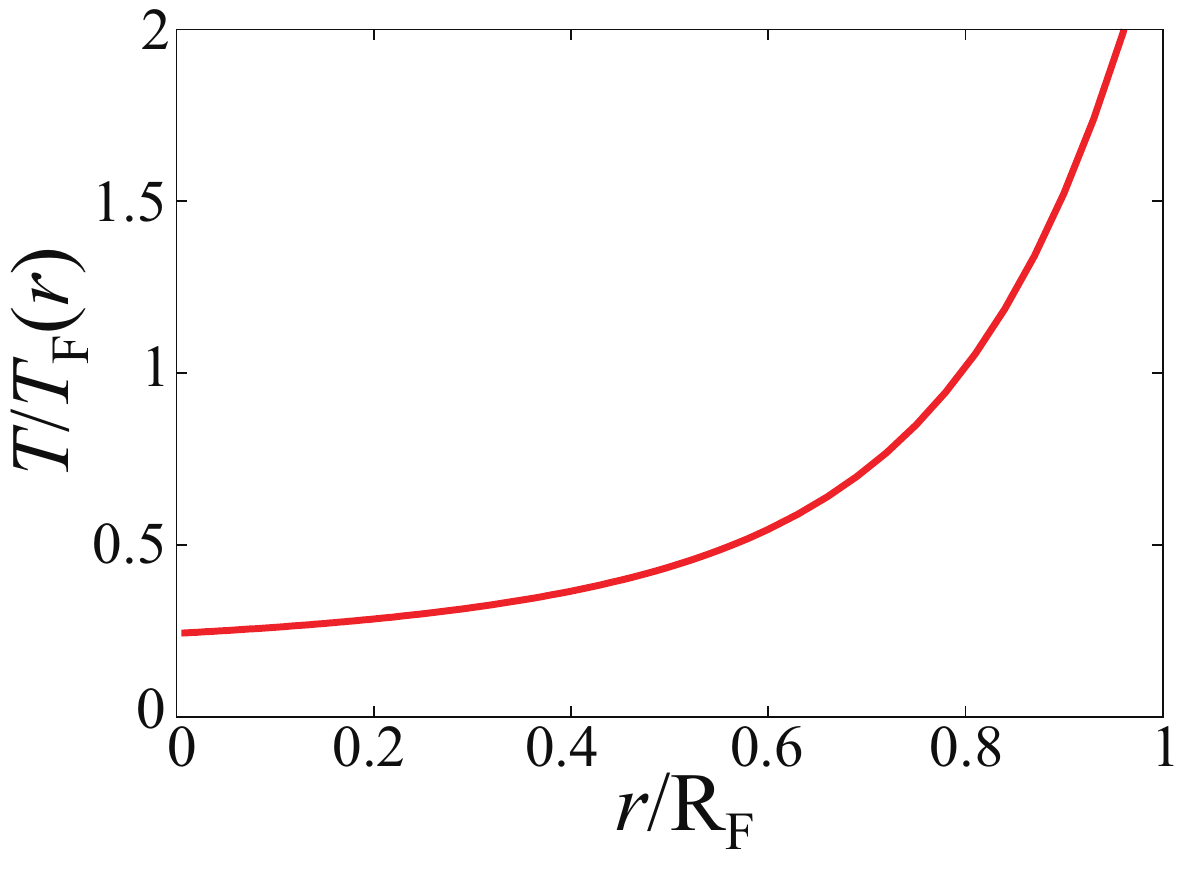}
\caption{Scaled local temperature $T/T_{\rm F} ({\bm r})$ at the unitary $(k_{\rm F}a_s)^{-1}=0$.} 
\label{Fig8}
\end{center}
\end{figure}
\par
The decrease of $h$ near $T_{\rm c}$ is related to the pseudogap phenomenon appearing in the central region of a trapped Fermi gas. To see this, it is convenient to introduce the LDA {\it local} Fermi temperature $T_{\rm F}({\bm r})$, given by,
\begin{equation}
T_{\rm F} ({\bm r}) = \varepsilon_{\rm F} - V({\bm r})~~~
(\varepsilon_{\rm F} \ge V({\bm r})).
\label{eq.25}
\end{equation}
Equation (\ref{eq.25}) is a natural extension of the ordinary Fermi temperature to the LDA case with the position-dependent Fermi chemical potential $\mu({\bm r})=\mu-V({\bm r})$. In Eq. (\ref{eq.25}), the Fermi energy $\varepsilon_{\rm F}=(3\pi^2n(0))^{2/3}/(2m)$ involves the particle density $n(0)$ of a trapped free Fermi gas at ${\bm r}={\bm 0}$. When the temperature is scaled as $T/T_{\rm F}({\bm r})$, it increases with moving away from the trap center, as shown in Fig. \ref{Fig8}. Because of this, even when strong pairing fluctuations cause pseudogapped single-particle excitations in the trap center below the so-called pseudogap temperature $T_{\rm pg}$\cite{Tsuchiya2011,note6}, pairing fluctuations are still weak in the outer region of the gas cloud where $T/T_{\rm F}({\bm r})>T_{\rm pg}/T_{\rm F}(0)$, leading to the vanishing pseudogap phenomenon there. This naturally gives the coexistence of the pseudogapped central region and the outer region with no pseudogapped local density of states.
\par
When $T_{\rm c}\le T\le T_{\rm pg}$, such spatially inhomogeneity would remarkably decrease $h$ from unity, unless the pseudogapped region spreads out over the entire spatial region where LPES observes atoms. In the case of Fig. \ref{Fig3} ($N_{\rm prob}/N=0.3$), the size of the spatial region where LPES observes atoms is as large as about half the Thomas-Fermi radius $R_{\rm F}$. We briefly note that, even at $T_{\rm c}$, the pseudogapped spatial region does not so spread out in the unitary regime\cite{Watanabe2012}. As a result, single-particle properties are still inhomogeneous in the region where LPES observes atoms, leading to the decrease of $h$.
\par
When $T>T_{\rm pg}$, the spatial inhomogeneity by the pseudogap phenomenon no longer exists. Because of this, a large value of $h~(\simeq 1)$ is expected compared to the pseudogapped case, which causes the difference between Figs. \ref{Fig6}(c) and (g).
\par
\begin{figure}[t]
\begin{center}
\includegraphics[width=0.6\textwidth]{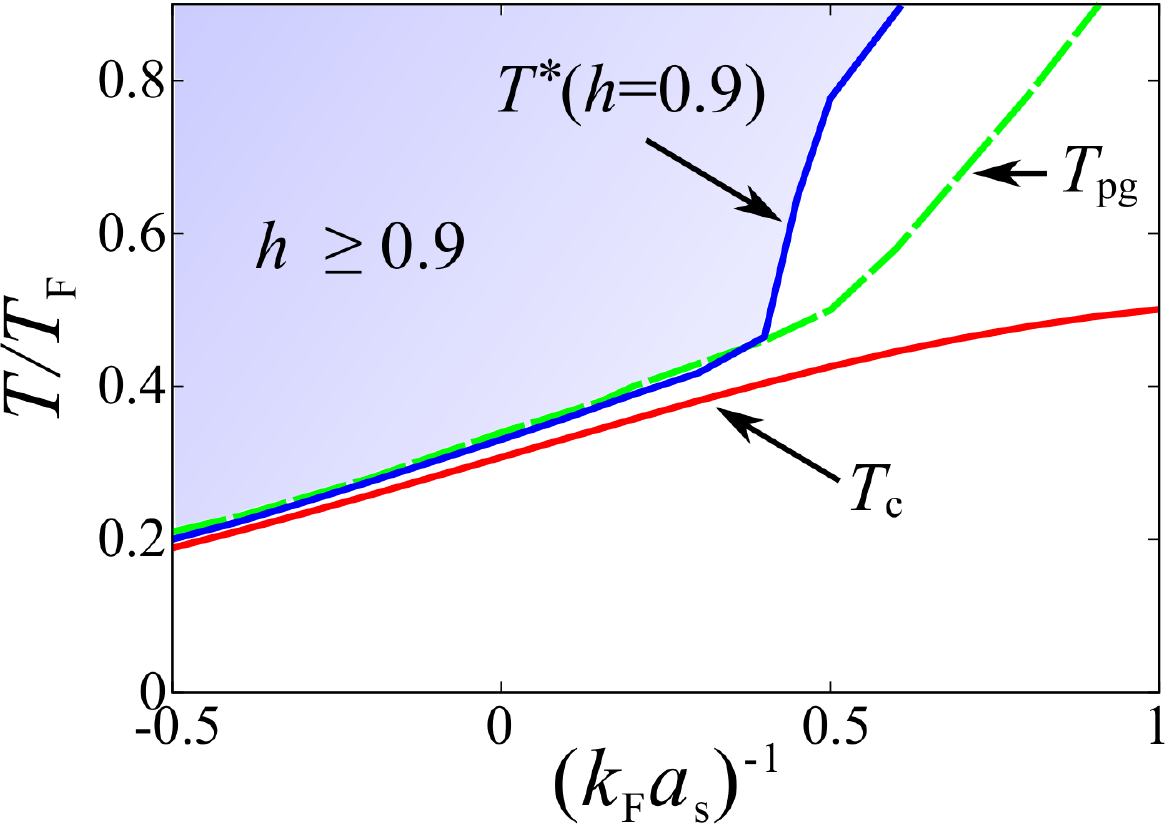}
\caption{Temperature $T^*$ above which $h\ge 0.9$ is realized when $N_{\rm prob}/N=0.3$, in the BCS-BEC crossover regime of a trapped ultracold Fermi gas. $T_{\rm pg}$ is the pseudogap temperature\cite{Tsuchiya2011} at which a dip structure (pseudogap) starts to appear in the single-particle density of state at the trap center. In the region $T_{\rm c}\le T\le T_{\rm pg}$, single-particle density of states around the trap center has a gap-like structure, in spite of the normal state.} 
\label{Fig9}
\end{center}
\end{figure}
\par
Although there is no clear boundary about the validity of the current LPES experiment (with $N_{\rm prob}/N\simeq 0.3$), it is convenient to introduce the quantity $T^*(N_{\rm prob}/N=0.3)$ which is given as the temperature at which the $h$-factor reaches $h=0.9$ (see Fig. \ref{Fig7}(b)). As a criterion, we then identify the region above $T^*$ (where $h\ge 0.9$) as the region where LPES can (approximately) eliminate effects of a harmonic trap from photoemission spectra. 
\par
Using this criterion, we conveniently identify the region where the current LPES experiment ($N_{\rm prob}/N=0.3$) works, as shown in Fig. \ref{Fig9}. In this figure, we see that the most of the pseudogap regime ($T_{\rm c}\le T\le T_{\rm pg}$) is outside this region. Thus, the current LPES experiment with $N_{\rm prob}/N\gesim 0.3$ needs further improvement, in order to confirm the pseudogap phenomenon in the BCS-BEC crossover regime of a {\it uniform} Fermi gas. At present, it seems difficult to select narrower spatial region so as to give a larger $h$, because of the limitation of the detectable spectral intensity in the current experimental technology\cite{Sagi2015}. Thus, a promising idea is to combine the current LPES technique with the non-harmonic (box-type and cylindrical) trap potential\cite{Hadzibabic2013, Zwierlein2016} mentioned in Sec. I. Then, the spatial inhomogeneity inside the selected region would be suppressed to some extent.
\par
\section{Summary}
\par
To summarize, we have discussed single-particle excitations in the BCS-BEC crossover regime of a trapped ultracold Fermi gas. Including pairing fluctuations within a strong-coupling $T$-matrix approximation (TMA), as well as effects of trapping potential using the local density approximation (LDA), we have calculated the local photoemission spectrum $I_{\rm LPES}({\bm p},\omega)$ in the normal state above $T_{\rm c}$. We showed that our results agree well with the recent local photoemission spectroscopy (LPES) experiment on a $^{40}$K Fermi gas, without introducing any fitting parameter. To conveniently estimate the similarity between the observed photoemission spectrum in a trapped Fermi gas and that in the homogeneous case, we have introduced a quantity (which is referred as the $h$-factor in this paper), to show that the recently observed LPES spectra in the unitary regime of a trapped $^{40}$K Fermi gas are very close to those in a uniform Fermi gas. 
\par
However, we also found that the current LPES experiment (which requires the probing rate $N_{\rm prob}/N\gesim 0.3$ to obtain detectable spectral intensity) does not always work well, for the purpose of eliminating effects of a harmonic trap from the spectrum. In particular, we showed that the $h$-factor remarkably deviates from unity in the most of the pseudogap regime when $N_{\rm prob}/N\gesim 0.3$, which means that the current photoemission spectrum still involves effects of spatial inhomogeneity there. Thus, in order to use LPES to resolve the debate about the existence of the pseudogap in the BCS-BEC crossover regime of a uniform Fermi gas, it is necessary to improve this experiment beyond the current limitation ($N_{\rm prob}/N\gesim 0.3$).
\par
For this improvement, one needs to either observe a smaller spatial region around the trap center, or use another type of trap potential. In this regard, the recent experimental work on non-harmonic trap\cite{Hadzibabic2013, Zwierlein2016} is promising. Since it gives a flatter potential than the ordinary harmonic potential in the central region, the combined LPES with such a non-harmonic trap may enable us to obtain the photoemission spectrum with $h\simeq 1$ within the current experimental limitation ($N_{\rm prob}/N\gesim 0.3$). The theoretical confirmation of this expectation is our interesting future problem. Extracting homogeneous information about a strongly interacting Fermi gas from experiments on a trapped Fermi gas is an important issue in cold Fermi gas physics, especially when this highly tunable system is used as a quantum simulator for other {\it homogeneous} Fermi systems. Thus, our results would be useful when such an application is intended for the study of single-particle excitations.
\par
\par
\section*{Acknowledgement}
We acknowledge T. E. Drake for providing us the experimental data in Fig. \ref{Fig6}. This work was supported by KiPAS project in Keio University. D.I. was supported by Grant-in-aid for Scientific Research from JSPS in Japan (No.JP16K17773). Y.O. was supported by Grant-in-aid for Scientific Research from JSPS in Japan (No.JP15H00840, No.JP15K00178, No.JP16K05503).
\par
\par


\begin{thebibliography}{99}
\bibitem{note1} We assume that $|\uparrow\rangle$ and $|3\rangle$ feel the same trap potential $V({\bm r})$, for simplicity.
\bibitem{note2} In LDA, the case of an anisotropic trap $V({\bm r})=m\sum_{i=x,y,z}\omega_i^2r_i^2/2$ can be always transformed to the isotropic case, by simply rescaling the spatial variables as $r_i\to (\omega_{\rm tr}/\omega_i)r_i$. Thus, we only deal with the isotropic case in this paper.
\end{thebibliography}
\end{document}